\documentclass{amsart}
\usepackage{graphicx}
\theoremstyle{definition}

\theoremstyle{remark}

\numberwithin{equation}{section}
\begin{document}

\title[DRAFT]{A Random Structure for Optimum Cache Size Distributed hash table (DHT) Peer-to-Peer design.}%
\author{Nima Sarshar and Vwani Roychowdhury}%
\address{Department of Electrical Engineering, University of California, Los Angeles, CA 90095}%
\email{{nima,vwani}@ee.ucla.edu}%

\keywords{peer-to-peer,content addressable,scalable,constant cache,Freenet}%

%\date{}%
%\dedicatory{}%
%\commby{}%
% ----------------------------------------------------------------
\begin{abstract}
We propose a new and easily-realizable distributed hash table
(DHT) peer-to-peer structure, incorporating a random caching
strategy that allows for {\em polylogarithmic search time} while
having only a {\em constant cache} size. We also show that a very
large class of deterministic caching strategies, which covers
almost all previously proposed DHT systems, can not achieve
polylog search time with constant cache size. In general, the new
scheme is the first known DHT structure with the following
highly-desired properties: (a) Random caching strategy with
constant cache size; (b) Average search time of $O(log^{2}(N))$;
(c) Guaranteed search time of $O(log^{3}(N))$; (d) Truly local
cache dynamics with constant overhead for node deletions and
additions; (e) Self-organization from any initial network state
towards the desired structure; and (f) Allows a seamless means for
various trade-offs, e.g., search speed or anonymity at the expense
of larger cache size. \iffalse Hence, unless totally new design
paradigms are invented, deterministic caching strategies do not
seem to be able to
outperform this design.\\
Constant cache size, allows for constant overhead for entering and
departure of nodes into the network. No previously known DHT
structure can achieve this with less than $\Theta(log(N))$
overhead. \\
The randomized nature of the design will allow for a high degree
of flexibility and dynamics, making the network very hard to track
and attack. The caching scheme is totally decentralized and does
not depend much on the availability of certain nodes. A totally
distributed algorithm for continuous update of cache contents are
also provided and the global convergence of any initial network
towards the desired structure is proved.\\
 Other generalizations
and modifications for trading off the performances is also
provided. \fi
\end{abstract}
\maketitle
% ----------------------------------------------------------------
\section{Introduction and Motivation }
 In general, the structure
of a DHT peer-to-peer network can be modeled as follows:
Each content when introduced into the network is assigned a key.
The key is uniformly chosen from a numerable set $R$ containing $N$ all possible keys.
This set can be a subset of the $d$ dimensional hypercube $Z^{d}$ as in \cite{Karp} or simply the one dimensional set $1,...,N$.
 Suppose for a moment that each node $I$ can provide the content $K_{I}$ when it is asked to.
 This assumption can easily be relaxed in future, but for now, it keeps the arguments simpler.
 Henceforth, we interchangeably use the key $K_{I}$ to refer to the physical node $I$ when there is no ambiguity.\\
 Note that the fact that the node $I$ has content $K_{I}$ does not really mean it actually has cached the file $K_{I}$, it instead might refer the search to a place where the actual file can be downloaded. Nevertheless, we assume that the search for content $K_{I}$ is hit the answer when reached the node $I$.\\
The node $I$, besides having the content $K_{I}$, has the address of $d$ other nodes $C_{K}=\{{K_{1},...,K_{d}}\}$ called its cache content.\\
A caching scheme is a deterministic or randomized rule that determines the subset $C_{K}$ of $R$ that the node $K$ has to cache in order for the final searching scheme to be able to find the queries initiated by arbitrary nodes for arbitrary keys.\\
This scenario as explained later,contains almost all proposed
peer-to-peer structures so far.

 \subsection{Deterministic DHT Schemes:}  By a deterministic caching scheme,
 we mean a deterministic algorithm that determines
 which keys should be addressed in the key cache of the node $K$ based on $K$.
 In other words, the topological properties of the emerging network directly specifies
the ability of the network to find queries for keys initiated at
different nodes.
 Suppose $R$ is supplied with an addition operator $+$.
  We define a commuting deterministic caching strategy as a set  of $d$ different elements of $R$, to be called $c_{1},...,c_{d}$,
  such that the cache of the node $K$ is $C_{K}=\{K+c_{1},…,K+c_{d}\}$.
   Note that although restricted at first glance, this definition includes almost all
   non-hierarchical structures proposed so far, such as Chord and CAD.\\
    For Chord \cite{chord}, the keys
are one dimensional on a ring lattice.
 Each node has $log(N)$ references to nodes with keys $N/2,N/4,N/8,…,2,1$ apart.
 Apparently, such a network can be searched in time $\Theta(log(N))$.
 When a node $K$ starts a query to find the key $T$, it passes the query to any node in its cache table which is closer to the target.
 Proceeding this way, it is not hard to see that any query  will receive the target in $O(log(N))$ steps.\\
The content space of CAD \cite{Karp} , on the other hand has a $d$
dimensional hypercube topology. Any node has in its cache the
address of its $2d$ neighbors.
 For $d=log(N)$ it is easy to show that the search time is
 $O(log(N))$.\\\\
\textbf{Result 1:} For a commutable deterministic caching scheme
as above, suppose $d\leq
\frac{log(N)}{mlog(log(N))}(1+log(\alpha)/log(N))$, then starting
at each arbitrary node $K$, at most $\alpha N$ of all
nodes can be reached in less than $log^{m}(N)$ steps.\\\\
\textbf{Proof:} Starting from node $K$, lets count the number of
different nodes that can be reached in at most $log^{m}(N)$ steps.
At each node $K'$, the query can proceed any of the $d$ different
links taking it to nodes $ K'+c_{1},...,K'+c_{d}$. After $l$
steps, the position of the query is $ K+a_{1}c_{1}+...+a_{d}c_{d}$
and $\sum_{i=1}^{d}|a_{i}|\leq l \leq log^{m}(N)$ hence
$a_{I}<log^{m}(N)$. The different number of the final locations
possible after $log^{m}(N)$ steps is bounded by:
$T=(log^{m}(N))^{d}$. Equating this by $\alpha N$ and taking the
logarithm from both sides we get: $d=
\frac{log(N)}{mlog(log(N))}(1+log(\alpha)/log(N))$. Hence, for $T$
less than this amount, at most the total number of all possible
nodes the query can reach in $log^{m}(N)$ steps is $\alpha N$.
There is no deterministic caching scheme achieving an average search time of polylog(N) unless $d=\Theta(\frac{log(N)}{log(log(N))})$.\\\\
\textbf{Corollary: }There is no commuting deterministic caching
strategy that allows for local or global search in mean polylog
time when the cache size is less than $\Theta(log(N)/log(log(N))$.
 Specifically there is no commutable deterministic caching scheme with constant cache size that allows for polylog search.\\\\
\textit{Result 1} can be easily extended to any finite commutable
set of operations rather than $+$. Note that in the proof, we only
assumed that the set of operators each node can choose from, is
finite and independent of the node position $K$.
 Note that a tree structure can be fit to this general scheme, however, each node chooses a constant number of operators from the total $log(N)$
 number of operators. Hence, a tree can achieve logarithmic search time with constant cache size.
 However, tree suffers from a serious problem, making it an improper structure for a p-2-p system.
  Sine the cache selection rule depends on the key $K$, the structure is not symmetric and hence the traffic is not distributed evenly,
   in fact deleting just one node might completely destroy a big
  number of routings paths.\\

 \section{Freenet: An attempt for a less structured design}
 In Freenet \cite{Freenet}, the topology is a ring
lattice. At the steady state, each node is meant to be connected
to $L$ nearby neighbors as well  as a few far neighbors called
shortcuts. There is no specific caching scheme as to what the far
keys cached should be. Freenet is important for our discussion
because it is among the few less structured designs that as will
be shown shortly has many
features in common with our proposed design.\\
The original Freenet design does not seem to be aware of the
crucial importance of random shortcuts for the scalability of the
system. There is  no simulation regarding the scaling behavior of
Freenet as a function of the cache size in hand, however the
original simulations of Freenet show that a relatively large cache
size of about $250$ will fail to provide reasonable search time
when there
are almost $200000$ contents in the network.\\
Based on the small world intuition, that is trying to have a
cluster of keys around a single key and having random shortcuts
apart from that key, modifications to Freenet have been proposed
in \cite{Berk}. They also conclude that $O(log^{2}(N))$ cache size
is necessary to perform search in time $O(log^{2}(N))$ on their
modified Freenet.\\
The problem with the Freenet is that, when pointed as the source
of a key by another node, a node might not in fact have that key
as the center of its cluster and hence might not be able to
provide a suitable search path for the query.\\
We will  show that this in fact might make the Freenet incapable of scaling as seems to be the case with the current Freenet.\\
 \textbf{\textit{Freenet is not scalable:}} Freenet, or at least with its
modifications in \cite{Berk}, apart from its efforts to enforce
anonymity, reads as follows: Each node has a seed key $K$. This
node might be called node $K$. When being in the routing path for
a key $K'$, the key $K'$ once found is cached if it is closer to
the seed than another key already in the seed. The variation to
allow for the shortcuts, allows a far key to replace a closer key
with some probability $q$. The intuition behind this is to have an
emerging ring lattice network topology, where every node is
connected to
$C$ neighboring nodes, while it has $qC$ shortcut links.\\
There are variations to how the Freenet updates its cache contents
mostly based on  different engineering intuitions. As an example,
the least referenced key might be substituted with a new key.
There however is no reason why any such caching strategy
should lead to a scalable search algorithm.\\
Approximating the emergent structure of the Freenet, with a local
connection of $2C$ links to $C$ neighbors on each side along with
$C'$ uniformly randomly chosen shortcuts we have the following
result:\\\\
\textbf{Result 2:} The Freenet with above  assumptions can perform
search in time $\leq log^{m}(N)$ iff
$(C+C')=\Theta(\frac{N^{1/2}}{log^{m/2}(N)})$.\\ More precisely,
take $\varepsilon =\frac{(C+C')}{2\frac{N^{1/2}}{log^{m/2}(N)}}$,
then for taking $C=C'$, the probability of not finding the target
in $log^{m}(N)$ steps is at most $e^{-2\varepsilon^{2}}$. On the
other hand for $\varepsilon<<1$ regardless of the choice of $C$
and $C'$, the probability of finding the target in $log^{m}(N)$
steps is at most $\varepsilon^{2}(1+o(1))$.
\\\\
\textbf{Proof:} We first prove the if part: Starting from any
node, consider a region of size $2Clog^{m}(N)$ with the target in
the middle. At each step of the walk the probability that there is
no shortcut into this region is
$p=(1-\frac{2Clog^{m}(N)}{N})^{C'}$ because there are $C'$
shortcuts at each node visited. Once a shortcut is found into this
region, then the target is simply reached in at most the next
$log^{m}(N)$ steps by the local connections.This probability is
bounded by $p\leq e^{-2CC'log^{m}(N)/N}$, now if
$CC'=\varepsilon^{2}N/(2log^{m}(N))$ the probability of not
fining the proper link after $log^{m}(N)$ steps is bounded by:\\
$p\{no\quad link\quad in\quad log^{m}\quad
steps\}=(1-2C/N)^{log^{m}(N)C'}<e^{-2CC'log^{m}(N)}=e^{-2\varepsilon^{2}}$.\\
Of course for this choice of $CC'$ the minimum cache size $C+C'$
is when the two are equal , that is
$C=C'=\varepsilon(\frac{N^{1/2}}{log^{m/2}(N)})$ completing the first part of the proof.\\
To see that this is in fact also necessary, consider the contrary.
\\ Consider two sets of nodes that are
more than $log^{m}(N)C$ apart. \\
Hence to reach the target in
$log^{m}(N)$ steps, at least one shortcut should be made into the
regions of width $C$ around the target. In the greedy search
followed by Freenet,the query gets closer to the target at each
step, however, since the shortcuts are uniform, this does not
change the probability of finding a shortcut into the specified
region. Hence in any case, the probability of finding a shortcut
into that region in the first $L$ steps is:$p=1-(1-2C/N)^{LC'}$.
The product $\alpha=2CC'L/N$ is bounded by the choice of
$C=C'=\varepsilon(\frac{N^{1/2}}{log^{m/2}(N)})$ giving
$\alpha<\varepsilon^{2}$ Expanding $p$ noting that $\alpha$
is very small:\\
$p=1-(1-\alpha+O(\alpha^{2}))=\varepsilon^{2}(1+o(1))$ completing
the proof.$\square$
\\\\
Hence unless a precise caching policy is not employed the Freenet
does not seem to be capable of scaling with ad hoc caching schemes
currently in use.\\
 In fact the necessity of a cache size of the
order of $O(N^{1/2})$ for a fast search can be tracked in the
early simulations of the Freenet \cite{Freenet} in which case a
cache size of 250 was used. The search time shows exponentially
abrupt increase as the system
size approaches $40000$ with $5$ keys per node. Note that $\sqrt{100000}\approx 300$.\\\\
The importance of caching strategy is hence evident. We propose a
practical peer-to-peer system based on precise reconsideration of
a rather similar architecture. \\\\
\section{A truly p-2-p random DHT structure}
With discussions in previous sections, new paradigms for designing
DHT peer-to-peer structures turn out to be necessary.\\
A p-2-p system (at least with homogenous members) should have the
following characteristics:\\
\textbf{1) Short routing paths:} This seems to be the first
desired characteristics of any p-2-p structure. A scalable design
is the one with path length scaling logarithmically with the
network size. From the four well known DHT designs, CHORD,
TAPESTRY and PASTRY, have search time $O(log(N))$, while CAN has
search time $O(dN^{1/d})$ for a constant $d$.\\
\textbf{ 2) Small Number of Neighbors or Cache Size:} The number
of neighbors of a node, meaning the number of keys it has the
address of, is probably the second important issue in a DHT
design. Particulary  there are two issues in having large cache
sizes. First the mere notion of space complexity which is more
relevant to theoretical issues than practical ones because of the
relative small
price of storage capacity. \\
The more important issue is the overhead involved in updating
these caches. P-2-P networks are very dynamic meaning that nodes
continuously join and leave the network. In fact one might assume
the query rate to be in the order of log-in rate, meaning that
many nodes just log-in to make constant number of queries and then
they log-off. Hence a small cache size is very important. The only
structure with constant cache size known so far is CAN.
Unfortunately CAN can not have polylogarithmic search time while
still having a constant cache size.\\
\textbf{3) Uniform Load Balance:} The load of query routing should
be as evenly spread as possible. At the very least, when queries
are made from unifomly random nodes to uniformly random targets,
the average load on all nodes should be equal. This (as well as
many other reasons) excludes tree like structures as candidates
for a DHT
structure.\\
\textbf{4) Patternless structure:} A deterministic design has a
very distinct connection pattern between its nodes. An attacker
can simply follow the pattern to disable a certain node. Suppose
as an example the CHORD design. An attacker now in node $K$, knows
exactly the set of keys $K$ should have in its cache. It then only
suffices to disable those keys for the node $K$ to be once and for
all excluded from the system. Less patterns in the connections
makes it less likely for an attacker to systematically attack the
network. All deterministic systems can be easily disabled by a
smart enough distributed attack.\\
\textbf{5) Locally adjustable tradeoffs:} A good p-2-p protocol
should consider the fact that the same algorithm might need to be
implemented in different working environments composing of
potentially different characterizations. As an example, it might
be desirable that the same system work in different problem
size regimes or different dynamical situations. \\
For instance, if the system dynamics is slow, one might want to
decrease the search time by using larger cache size by local
adjustments to the protocol. Another issue is security and
anonymity. It might be desirable to have a structure that can
switch between an anonymous mode of operation and a normal one, by
trading off different system specifications to some extent.
\\\\
With these goals in mind, we propose the first known randomized
DHT structure with the following characteristics:\\
1) Randomized caching strategy with constant cache size O(1).\\
2) Average search time of $O(log^{2}(N))$.\\
3) Guaranteed search time of $O(log^{3}(N))$. \\
4) Truly local cache dynamics with no overhead.\\
5) Global convergence from any initial network state.\\
6) Local adjustments can trade cache size for search speed.\\
7) Anonymity can be bought in price of larger cache size with only
local considerations.\\\\

The following lemma due to Kleinberg is essential to our design
structure:\\
 \textbf{Lemma 1} (Kleinberg) \cite{Klein}: Consider a lattice topology in $Z^{d}$. Nodes are
placed on the grid points. Each node is connected to all its $2d$
neighbors. Also each node has a shortcut link. The probability
that the node $K$ has a shortcut to node $K^{\prime}$ is
proportional to the inverse of its $d$ dimensional Euclidean
distance, that is $p(c_{K}=K^{\prime})\propto
1/|K-K^{\prime}|^{d}$. A greedy search algorithm starting from a
random node searching for a random target, will find the target in
average time $O(log^{2}(N))$. Also, for no other exponent in the
probability rather than $d$ the average
search will have polylogarithmic time.\\\\
\textbf{Proof:} Please refer to \cite{Klein} for a complete proof of the case $d=2$. For the case $d=1$ however we sketch the proof as it proves relevant to our other considerations in the rest of the paper:\\
Divide the region from $1,...,N$ into $log(N)$ distinct regions
$X_{i}=\{2^{i},...,2^{i+1}-1\}$. Suppose the target and the
current node possessing the query are a distance $r\in X_{i}$
apart. Then we say that the search in phase $i$. Each node
receiving the query, will pass it to a node in its cache which is
closest to the target. \\
The proof relies on the fact that at most
$O(log(N))$ steps are required for the query in each phase $i$ to
reduce its phase by at least one. To see why, note the furthest
distance from nodes in the
 phase $i$ to those in phase $i-1$ is $2^{i}-2^{i-1}=2^{i-1}$.
 Also the normalizing constant is $c log(N)$ for a proper $c$ meaning that the probability that a node in phase $i$
 has a link to a node in phase $i-1$ is at least
 $c2^{-(i-1)}/log(N)$.\\
 There are then at least $2^{i-1}$ nodes in phase $i-1$, hence the total probability that a node in phase $i$ has a
 link to the phase $i-1$ or less is at least $c'/log(N)$.\\
Now the probability that in $L$ steps in phase $i$ no link is
found to phase $i-1$ is at most $p=(1-c/log(N))^{L}$.
  Hence in average $O(log(N))$ steps a link is found to the next phase. Since there are $log(N)$ phases altogether,
   the query reaches phase $1$ which has $log(N)$ members in $O(log^{2}(N))$ steps. The last phase can be traversed by only local links in $log(N)$ steps at most.\\
   \\
\textbf{Result 3:} For any realization of the above scheme, there is a constant $a$ independent of $N$,
such that any query almost surely reaches any target at time at most $alog^{3}(N)$ as $N\rightarrow \infty$.\\\\
\textbf{Proof:} Considering the above proof, the probability of
not finding any link into the next phase after $3log(N)^{2}/c$
steps, for the $c$ defined in the Lemma 1, is bounded by $N^{-3}$,
there are $N^{2}$ possible queries making the probability of any
query having that problem arbitrary small. So each query will
reach the destination almost surely in at most $(3/c)log^{3}(N)$.
$\square$.\\\\
\\ Consider a space of at most $N$ node-keys as
described earlier formed as a ring lattice topology. Hence each
node has the
 address of two close nodes in its cache. Also each node $K$ has the address of another node $K'$.
 The probability of the node $K'$ to be chosen is proportional to $\frac{1}{|K-K'|}$. Kleinberg's
 theorem can be readily applied to this case to show that a greedy algorithm can search this network
 with time $O(log^{2}(N))$.\\\\
 \textbf{Result 4 :} There is a randomized caching strategy along with a local search that
 allows for average $O(log^{2}(N))$ search time, while each node has only $3$
 neighbors.\\\\
 \textbf{Proof:} Consider a DHT system employing the above caching strategy.
 That is a node having key $K$, also has the keys $K+1$ and $K-1$
 and another key $K'$ chosen with probability $\propto
 \frac{1}{|K-K'|}$ among all other keys
 (all operations of mod N). One can easily verify that the ring
 lattice topology does not alter the above lemma much.\\
 The number of neighbors of any node is hence just $3$. Lemma
 1, shows that such a network topology equipped with a greedy
 search algorithm has average search time of $O(log^{2}(N))$ and
 Result 3also predicts a worse case search time of $O(log^{3}(N))$.
 Hence with constant cache size, the search time is polylog,
 something proven impossible for at least a very large class of deterministic caching
 strategies.\\\\
Practically the above design will have the following
specifications:
\\\\
\textbf{1) Topology:} Each content is assigned a key uniformly
randomly chosen from the set of all possible keys $1,...,M$. This
assumption might restrict the future expansion ability of the
system. Nevertheless, choosing a very big $M$ does not add much to
the computational
costs of the system. In fact assigning $32$ bit integers to the keys,
naturally confines any design to a few billion keys.\\\\
At a particular time, there are $N$ nodes in the network. $N$
might be very smaller than $M$, however, the $N$ keys are assumed
to be uniformly distributed in the $M$ possible places. The
topology is then that of a ring with $M$ places where $N$ of them
are occupied. Note that a physical node, might in fact contain
more than one key. The important point is that for each key
introduced, a proper shortcut as well as close keys should also be
added to the cache contents. Different shortcuts due to different
keys of one node, need not be differentiated, since following a
closer shortcut can
only facilitate the search procedure.\\\\
\textbf{2) Node Arrivals:} When a node with a certain key joins
the network, it initiates a query for its own key to a bootstrap
node through which it has joined the network. The query passes
until reaches a node where it cannot pass the query to any node
with a closer key. This node, has the closest key to the new key.
The new node then makes two connections to this node, and a node
immediately after that in order for its local connections.\\
Now, to make its shortcut, the node pretends as if the key space
was not sparse. Node $K$ chooses a key $K'$ with probability
$|K-K'|^{-1}/H$, where $H=\sum_{i=1}^{M}1/i\sim log(M)$. It then
initiates a query for this key. Again, the closest key to this
query is returned. The
node then adds this query to its cache table. \\
Looking at the procedure of the Lemma 1, it is easy to see that
this way the sparse nature of the key space does not prevent the
efficient search provided that the keys are uniformly distributed.
In fact, every length scale, is scaled by a factor $M/N$ which
does not affect the analysis. In later sections, we will introduce
a more natural and practical self-organizing algorithm, which
allows for cache update as the network answers queries. This
eliminates the need for making dummy
queries for keys that might not exist, also does not require knowing $M$.\\\\
\textbf{3) Node Departure:} Though the original scheme does not
require links to be bidirectional, there are certain practical
advantages in having a bi-directional link. The most important of
all is that when a node decides to log-off, it can inform the
nodes referring to it to look for another random key. A convenient
way is to refer to a nearby node. A better way is to simply redo
the procedure
done for the connection.In any case it is antural to assume that a node is informed
about the log-off of any of its connections.\\\\
\textbf{4) Refreshing:} Nodes can change their shortcuts
completely without previous notice. Since never in the proof of
Lemma 1, it is necessary for a node to be certain about where its
links are before it receives the query. In fact the principle of
differed decision is in the heart of the proof and nodes can
dynamically change their shortcut connectivity. This way the
traffic can be evenly distributed among all nodes avoiding hot
spots due to the nodes receiving very far links. It is a good idea
to have a refreshing rate in the order of the network dynamics
time scale
(the log-n log-off rate).\\\\

 \textbf{Trading cache size for speed:} In
this section we introduce the concept of nested shortcuts to allow
for faster search times when more than constant cache size can be
tolerated. The idea is to provide more than one level of shortcuts
by aggregation of close keys an relabelling them. As an example,
consider the following aggregation
process:\\
Assign each set of nodes $N^{1}_{i}={(i-1)N/log(N),iN/log(N)-1}$ a
new label $i$. Call $N^{1}$ the set
${N^{1}_{1},...,N^{1}_{log(N)}}$. For each set $N^{1}_{i}$,
proceed the same way and define the sets $N^{1,1},N^{1,2}...$ by
aggregating the keys of each set, into smaller sets. The number of
sets resulting from each division is the logarithm of the size of
the upper set. This will continue until the number of elements in
the lowest hierarchy is $1$ (or even $log(N)$). Now at each level
of the hierarchy, a node with label $i$ will have a link to
another node from the same hierarchy with label $j$ with
probability proportional to $|i-j|^{-1}$. The normalizing constant
is chosen according to the number of labels on that hierarchy.
Among all nodes with label $j$, node $i$ will choose one in random.\\\\
\textbf{Result 5:} A greedy search on a structure made by the
above procedure, results in search time $O(log(N)loglog(N))$ while
the cache size is
$S<log(N)/log(log(N))$.\\\\
\textbf{Proof:} Suppose for a moment that only the links of the
very first hierarchy existed. Having a small-world in the first
hierarchy, after at most $O(log^{2}(N_{1}))$ steps, the query and
the target will match on the first label, where $N_{1}$ is the
number of the labels in the first hierarchy. Note that having
other links can only speedup this process. From then on, the same
arguments can be made for other hierarchies. The number of labels
in each hierarchy is at most $log(N)$. Hence the total search time
is $O(log^{2}(log(N)))W$ where $W$ is the total number
of the labels.\\
Now lets bound $W$ the depth of the hierarchy which in turn is the
cache size necessary. To do this, start with the first hierarchy
having $log(N)$ labels and $N_{1}=N/log(N)$ members in each label.
Taking the $log$ results in $log(N_{1})=log(N)-log(log(N))$. Hence
the members of the second label have
$N{2}=\frac{N}{log(N)log(\frac{N}{log(N)})}$ members.
So:\\
\begin{eqnarray*}\label{100}
    log(N_{2})&=&\\
    &=&log(N)-log(log(N)(log(N)-log(log(N)))\\
    &=&log(N)-log(log(N))-log(log(N)-log(log(N)))\\
    &=&log(N)-log(log(N))-log(log(N)(1-\frac{loglog(N)}{log(N)})\\
    &=&log(N)-2log(log(N))-log(1-\frac{loglog(N)}{log(N)})\\
    &=&log(N)-2log(log(N))+o(1)
\end{eqnarray*}
Proceeding this way, for other steps, its not hard to see that:\\
\begin{equation}\label{101}
    log(N_{i})=log(N)-i\times loglog(N)+o(1)
\end{equation}.
Assuming $N_{i}>1$, that is $log(N_{i})>0$, it implies that:
\begin{equation}\label{102}
    i<log(N)/log(log(N))
\end{equation}
Meaning that in at most $W<log(N)/log(log(N))$, the number of
labels gets to $1$. $\square$\\
One can trade cache size for the same increase in the speed. This
generalization  might prove useful in applications where faster
than $log^{2}(N)$ search is necessary. The limit of $O(log(N))$
cache size and search time is the same as that can be achieved by
deterministic means as well. However, the randomized nature of the
caching allows for more robust and less likely to be attacked
networks. Also, all decisions are totally local, and at each level
there are very many nodes from which the shortcut can be chosen
(in fact in the $i$th hierarchy
almost $\frac{N}{log(N)^{i}}$ different nodes can be chosen. Though probably of less
practical importance, this scheme shows the ability of the design
to be adapted to different working regimes.\\\\\\
\textbf{ Multiple contents and Anonymity:} So far we assumed each
node has exactly one key, with which we named the node itself. As
might have become clear, this assumption can readily be relaxed.
Any physical node can have multiple content keys. Each content
however should have its neighbor keys as well as a shortcut key
chosen according to the proper distribution around that content.
\\
Any search for any key in this new system can only be faster than
the case where only one content was present.\\
On the other hand different contents need not really represent
totally unique keys. Same key might be cached in many nodes.
Turning to the idea used in Freenet to ensure a degree of
anonymity one can think of the following algorithm: \\
A key once found would be cached, as if it was their own content,
on all the nodes in the search path. Of course the node caching it
must also provide the nearby neighbors as well as the proper
shortcut for this new key. The search is really said to be
finished when such a key is found.\\
The anonymity is insured by increasing the average cache size of
the nodes.\\
\section{local cache updates and topology convergence}
In previous sections, it is assumed that nodes are able to make
shortcuts in the content space to keys that are a distance $r$
apart with probability $\propto 1/r$. We investigate the
possibility
of performing such task locally.\\
If the size of the key space is known a priory, meaning that the
maximum capacity of the network is fixed, then each node joining
the network can initiate a query for a random key it chooses from
the $1/r$ distribution around its own key. If that key does not
exist, the closest key is cached. To take care of new nodes
joining, this procedure can be repeated with an appropriate rate.
\\ The limitation for knowing $N$, comes from the fact that the
distribution $1/r$, is not normalizable for $N\rightarrow
\infty$.\\
In this section we examine a more natural approach.The network
gradually organizes itself to take the form of the desired
topology as more queries are being answered by the system. Our
idea is based on the assumption that the queries for a key $K$ are
initiated by randomly chosen nodes in the key space. A node $S$
starts a query for the target node $T$. The key $T$ can be a very
popular key who receives many requests in time, however, the nodes
initiating the request for that key can be assumed to be uniformly
chosen from the network.\\
Now, as the system works, a node $K$ answers a request form the
node $T$. Suppose $K$ already has key $L$ as its shortcut in the
cache. Upon answering the request of the node $T$, it replaces $L$
with $T$ with probability $\frac{|K-L|}{|K-L|+|K-T|}$. Otherwise
it keeps the key $L$. Similar idea is suggested in \cite{Berk} as
an intuitive answer. Here we prove the validity of this intuition by the following theorem:\\\\
\textbf{Result 6:} Repeating this procedure the probability of
having the key $K'$ in the cache tends to be $\propto
\frac{1}{|K-K'|}$.\\\\
\textbf{Proof:} We define a Markov chain as follows: If the
distance of the cache from the node's own key $K$ is $x$, we say
that the Markov chain is in state $x$.\\
A step is made when a new sample with distance $y$ is received.
Then the chain walks to state $y$ with probability $x/(x+y)$,
otherwise it stays in the same state. This procedure clearly
defines a Markov chain with $N$ states. For any finite number of
states $N$ a stationary probability distribution for the process
exists. Lets call the stationary probability of residing in state
$x$, $p_{x}$. We will find this stationary solution considering
the flow-in flow-out of an arbitrary state $x$. The flow coming
out of the state $x$, is the probability of being in $x$ (i.e.
$p_{x}$) times the probability of moving out of it in the next
step.\\
The flow into the state $x$ is the probability of moving to the
stet $x$ in the next move. Equating the two:
\begin{equation}\label{103}
    p_{x}\sum_{i=1,...,N}\frac{1}{N}\frac{x}{x+i}=\frac{1}{N}\sum_{j=1,...,N}p_{j}\frac{j}{x+j}
\end{equation}
Hence, evidently $p_{i}=(1/c)\times\frac{1}{i}$ satisfies the
equation for all $x$ where $c=\sum_{i=1}^{N}1/i$ is the
normalizing constant. Since the stationary solution is unique the
system always
converges to the proper distribution.\\
Meaning that no matter what the initial keys assigned to the nodes
during the bootstrap section are, the system eventually formes
towards the desired solution as soon as enough random queries are
made.\\
Another interesting practical feature of this scheme is that the
nodes are naturally informed from the entrance of new nodes, when
those nodes make queries. The new nodes are then automatically put
in the searching routes by being added to the cache of the target.
\\
\section{Simulations} 
\begin{figure}
\includegraphics[width=3in,height=3in]{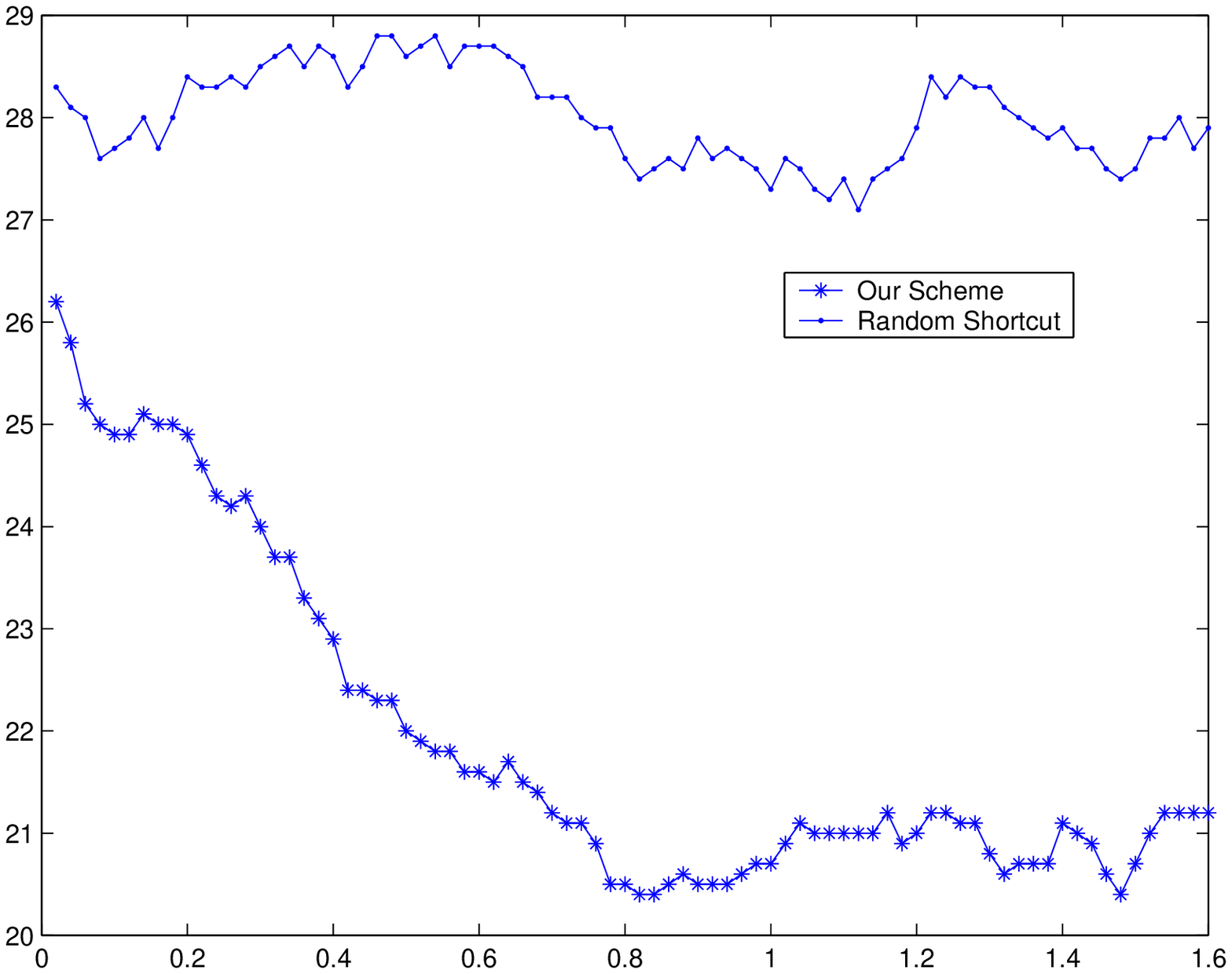}
\includegraphics[width=3in,height=3in]{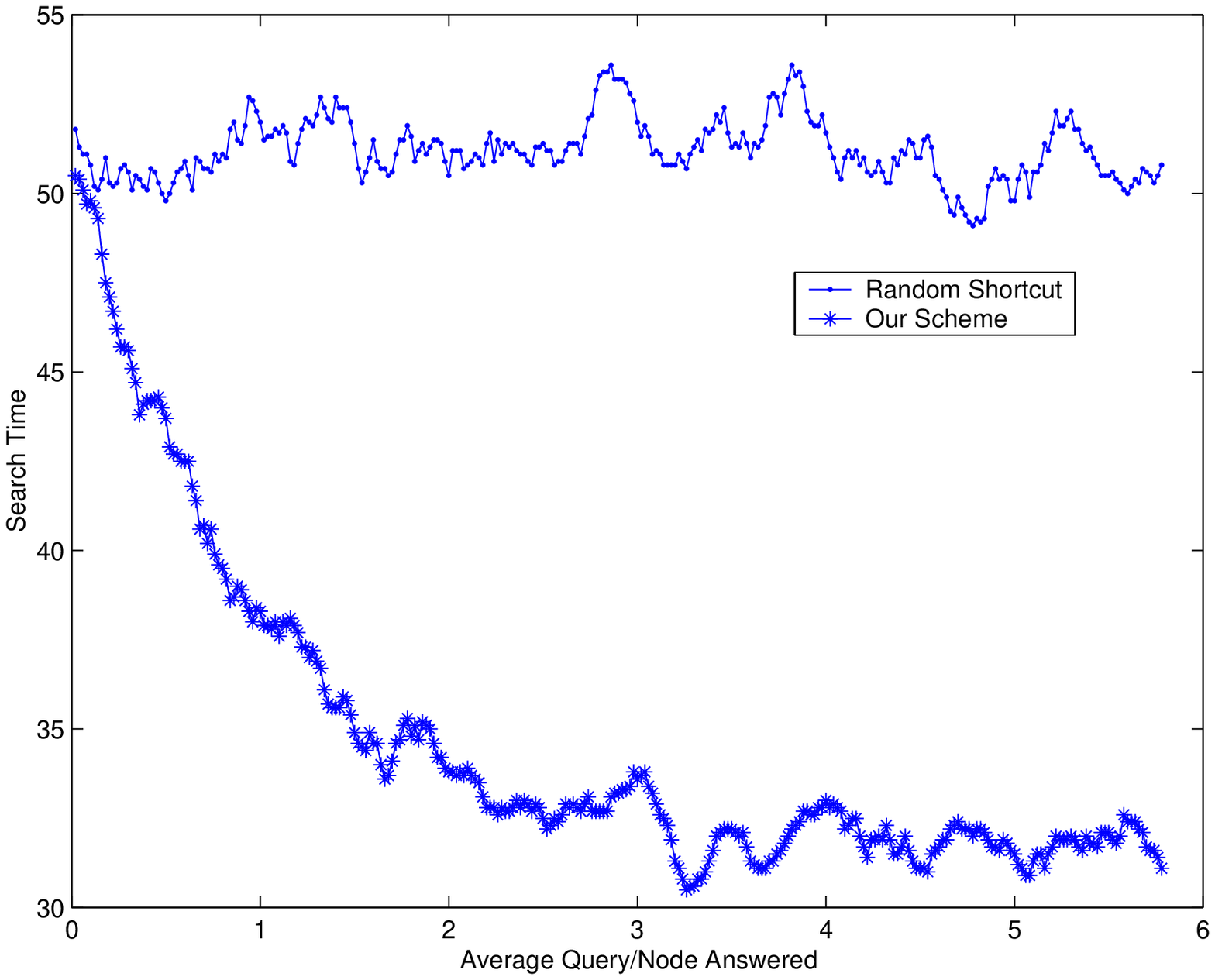}
  \caption{Average search time from the bootstrap phase until the system settles down to its
  steady state. The total number of nodes are $1000$ for the top figure and $3000$ for the bottom one.
  Included in each figure is the search time for a system with a random shortcut for comparison.The total key space is $10000$.
  Keys are uniformly randomly chosen. Each node( key) knows the address of the nearest neighbors as well as one shortcut.
  Queries are in randomly chosen pairs. The cache is updated through the rule in Result 6. As enough queries are made (about
  twice the total number of nodes) the system settles down. The average search time
  at steady state is around $21$ for $N=1000$ and $28$ for $N=3000$.}\label{search_time}
\end{figure}

To prove that the caching system in fact
can be used as a superior structure for a peer-to-peer network, we
have provided the following simulation. Starting form two nodes, a
node joins the network and is assigned a random key. It is then
placed between the two nearest keys already in the network,
meaning that it caches their address as the immediate neighbors,
the neighbors also update their caches to recognize the new
node.\\
It is also assigned a random key from the nodes previously in the
network as its shortcut. This key can be the node contacted by the
new node as the bootstrap. After $N$ nodes joined the network in
this fashion, at each time step two nodes are randomly chosen from
the network and one initiates a query for the other. Upon
reception of the query by the target, it replaces its cache with
the key of the requester with the mentioned probability. After
each time step, $100$ queries are made to random nodes in the
network to find the average search time.\\
The average search time is depicted in FIG.\ref{search_time}. As
clearly seen in the figure, the system totally reaches the steady
when the number of queries made is in the order of the number of
nodes meaning that almost all nodes have had the chance to update
their cache at least once. This is a totally satisfying result,
meaning that each node has to answer in the order of one queries
before
the network topology is formed. A precise formulation of the settling problem
is the subject of a subsequent paper.\\
\section{Conclusion} We showed that there is in fact little hope for
deterministic caching strategies in peer-to-peer networks to
provide logarithmic search time with constant cache sizes. Also
their deterministic structure makes them vulnerable to attacks.
Ourproposed system can provide search times in the order of
$O(log^{2}(N))$ with cache size of only $3$. We also showed that
the original Freenet lacks scaling, however our proposed strategy
provably provides scalable and highly efficient peer-to-peer
networks. It can also be used as a quick modification into Freenet
original
caching protocol.\\
Through out the derivations, routings to different nodes are
regardless of what shortcuts the receiving node has. This enables
nodes to change their shortcuts without prior notice. Hence even
during a single query search, the path might constantly change,
making it very hard to track the paths.\\
Also, the shortcuts are always being refreshed as new queries are
received by a node without producing any overhead traffic. The
symmetry of the protocol results in a naturally balanced load on
all nodes.\\
The randomized nature of the connections as well as their dynamic,
prevents an attacker from fragmenting the network.\\
Our future work is towards implementing a practical peer-to-peer
network based on the architecture proposed in this rather
theoretical paper by taking into account the more detailed
practical considerations such as duplicate keys and failure
handling.
\\
% ----------------------------------------------------------------

\end{document}